# Fostering Bilateral Patient-Clinician Engagement in Active Self-Tracking of Subjective Experience

**Jakob Eg Larsen**
Technical University of Denmark
and TOTTI Labs
jaeg@dtu.dk

**Thomas Blomseth Christiansen**
Konsulent Blomseth
and TOTTI Labs
thomas@blomseth.dk

**Kasper Eskelund**
Danish Defence Military Psychology Unit
VETC-MPA41@mil.dk

**Abstract**
In this position paper we describe select aspects of our experience with health-related self-tracking, the data generated, and processes surrounding those. In particular we focus on how bilateral patient-clinician engagement may be fostered by the combination of technology and method. We exemplify with a case study where a PTSD-suffering veteran has been self-tracking a specific symptom precursor. The availability of high-resolution self-tracking data on the occurrences of even a single symptom created new opportunities in the therapeutic process for identifying underlying triggers of symptoms. The patient was highly engaged in self-tracking and sharing the collected data. We suggest a key reason was the collaborative effort in defining the data collection protocol and discussion of the data. The therapist also engaged highly in the self-tracking data, as it supported the existing therapeutic process in reaching insights otherwise unobtainable.

**Introduction**
Based on our work with health-related self-tracking in various settings during the past decade we propose that the distinction between client-initiated and clinician-initiated self-tracking may be expanded and qualified further by considering both collaborative processes leading to self-tracking and specifically the decision-making regarding which phenomena could be subject to self-tracking. We surmise that bilateral engagement may be fostered through process focus and by technology with some—potentially—non-obvious characteristics.

Post-traumatic stress syndrome (PTSD) is a debilitating mental health disorder following upon exposure to different types of traumatic stress. The referenced case study [6] has its focus on military PTSD and bringing patient self-tracking data into the psychotherapeutic treatment. Additionally, in relation to our other experience we discuss some key challenges for using self-tracking data in clinical contexts.

**Case: PTSD Symptom Self-Tracking**
The PTSD patient (PN) is a Danish combat veteran suffering from severe depression, anxiety, and hyperarousal upon returning to civilian life after two deployments. While initial psychotherapeutic treatment had relieved depression and PN had learned strategies for reducing anxiety, PN was still suffering from seemingly unpredictable and uncontrollable peaks of arousal [6]. The therapist to a high degree depended on understanding the specific triggers of such episodes to target the treatment. However, cognitive alterations in PTSD may impede the patient's memory recall and therefore preclude the therapist from obtaining important details about the patient's condition.

The therapist offered the patient self-tracking as a voluntary option during the course of the treatment, thereby initiating the process that lead to the patient's personal self-tracking practice. No standard definitions of what to track in relation to PTSD were enforced upon the patient. Instead the patient and therapist during a two-hour assessment session jointly developed and agreed on the self-tracking of a single phenomenon – a bodily experienced precursor to hyperarousal. For recording observations of the precursor to hyperarousal our novel wearable self-tracking instrumentation consisting of a wrist-worn smartbutton connected via Bluetooth to a smartphone [6, 5] was used. Upon making an observation of the defined phenomenon the patient would promptly press the smartbutton to register the occurrence. This would leave a trace of individual observations of the precursor to hyperarousal. As the usual therapeutic insights are limited by the patient's memory and his willingness and ability to recall within therapy, the goal of self-tracking was to momentarily externalize occurrences of the precursor to hyperarousal as they unfold throughout daily life and thereby avoid the recall bias impacting interviews or questionnaires.

All observation data collected by the patient was persisted locally on the smartphone and belonged to the patient. As part of a therapeutic session the patient would have the choice to share data with the therapist as a high-resolution diary of symptoms. This left the patient in control of the data and of the decision whether to share the data. The shared data were a time series of observations that the therapist would plot to see the frequency and time of day of hyperarousal precursor occurrences. Particular data points were used as starting points for discussions of the triggers of individual symptoms, thus functioning as a memory aid for the patient to recall situations from memory that otherwise might not have surfaced during ordinary therapeutic sessions. An overview of the observations made by the patient over a 100-day period is shown in Figure 1 (see [6] for further description).

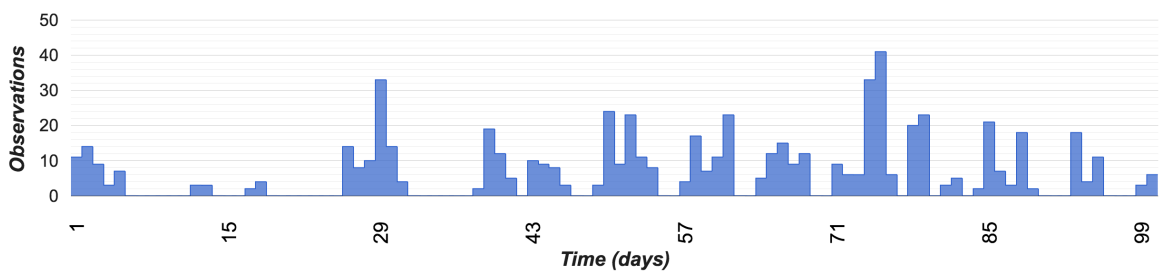

Figure 1: Number of observations made per day by the patient over a 100-day period [6].

**Discussion**
We deem it of paramount significance that the decision on using self-tracking and how to specifically use it was made in a dialogue between patient and therapist. As mentioned no predefined phenomena of interest or standard one-size-fits-all procedure were applied. Instead self-tracking was tailor-made in collaboration with the particular patient for the particular situation.
The patient engaged actively and persistently in longitudinal self-tracking (see Figure 1). We suggest that a key factor leading to the high level of engagement was the initial dialogue about the self-tracking method and protocol. We propose that the combination of method and technology lead to a very concrete form of the often elusive concept of patient empowerment: Every observation made by the patient was about something the patient cared about and found relevant. Regarding the therapist's engagement in the self-tracking process we suggest that it was to a large extent due to the self-tracking data fitting well into an already existing flow in the psychotherapeutic process and the self-tracking data providing access to information that was otherwise difficult or impossible to get access to through interviews or questionnaires due to memory recall bias.

Thus a bilateral patient-therapist engagement was facilitated through the use of the self-tracking technology. The data collected by the patient created a third entity in the therapeutic space enabling patient and therapist to discuss triggers of symptoms and identify causes in a much more qualified way than previously. We believe this transcends "improving decision making" and suggests that self-tracking data on a fundamental level may change what we know about health—and ourselves.

The referenced case study exemplifies the use of self-tracking as part of psychotherapy, whereas in other clinical settings where we have worked on integrating self-tracking (autoimmune diseases, urology, heart failure) we have found serious barriers to adoption. In current medical practices these types of data are not a first-class citizen in clinical decision-making. It appears as if medicine presently does not have the mental models and methods to learn from such data, why it seems as there is a serious epistemological gap needing to be bridged for this to happen.

**About the Authors**
The authors will contribute with substantial experience in self-tracking and from the Quantified Self (QS) Community. Besides having participated in all QS conferences we have participated in numerous US and EU meetups. JEL has founded QS Copenhagen.

TBC has more than 8 years of experience working with self-tracking technology and methods [12]. Through longitudinal active self-tracking of a number of different phenomena he has fixed his pollen allergy to an extent where he has not used any allergy medication for almost 7 years [1]. His unique data set includes e.g. 5 years of sneezes that have been instrumental in obtaining new insights and likely is one of the world's largest active self-tracking data sets [11]. Previously (2009-2015) he was a co-founder of Mymee, a service enabling clients and health practitioners to use active self-tracking for alleviating idiopathic health conditions.

JEL is Associate Professor at Technical University of Denmark where he is heading the Mobile Informatics and Personal Data Lab researching the phenomena of self-tracking and quantified self. He has been developing and researching personal informatics systems for different health purposes and has co-organized several workshops on the topic at CHI [8, 7, 3] and UbiComp [9, 10]. Moreover he has substantial personal self-tracking experience.

KE holds a PhD from the Technical University of Den–mark. He is working as a researcher and clinical psychologist at the Danish Defense with research and treatment of service personnel suffering from PTSD. This involves psychotherapy, EEG neurofeedback and the use of EEG to study correlations between self-reported symptoms and alterations of neural response patterns [2, 4].